\documentclass[aps,prd,
twocolumn,
preprintnumbers,groupedaddress]{revtex4}
\usepackage{epsbox}
\usepackage{amsmath}
\usepackage[dvips]{graphicx}

\begin{document}
\preprint{\vtop{{\hbox{YITP-06-49}\vskip-0pt
                 \hbox{KANAZAWA-06-11} \vskip-0pt
}}}

\vspace{20mm}

\title{$\bf{D_{s0}^+(2317)}$ as an iso-triplet four-quark 
meson\footnote{Invited talk at 
{\it QNP06}, the 4th International Conference on Quarks and Nuclear 
Physics, June 5 -- 10, 2006, Madrid, Spain}
}
\author{Kunihiko Terasaki}
\affiliation{Yukawa Institute for Theoretical Physics, Kyoto University,
Kyoto 606-8502, Japan\\
Institute for Theoretical Physics, Kanazawa University, 
Kanazawa 920-1192, Japan}
\thispagestyle{empty}

\begin{abstract}
Although assigning $D_{s0}^+(2317)$ to the $I_3=0$ 
component $\hat F_I^+$ of iso-triplet four-quark mesons is favored 
by experiments, its neutral and doubly charged partners have not yet 
been observed. It is discussed why they were not observed in 
inclusive $e^+e^-\rightarrow c\bar c$ experiment and that they can 
be observed in $B$ decays. 
\end{abstract}
\maketitle

The charm-strange scalar meson $D_{s0}^+(2317)$ has been observed in 
inclusive $e^+e^-$ annihilation~\cite{BABAR-D_{s0},CLEO-D_{s0}}. 
It is very narrow ($\Gamma < 3.8$ MeV~\cite{BABAR-search}) and it 
decays dominantly into $D_s^+\pi^0$ while no signal of 
$D_s^{*+}\gamma$ decay has been observed. Therefore, the CLEO 
provided a severe constraint~\cite{CLEO-D_{s0}}, 
\begin{equation}
R(D_{s0}^+(2317))    
< 0.059,                                     \label{eq:constraint}
\end{equation}
where 
$R(S)= {\Gamma(S\rightarrow D_s^{*+}\gamma)}/
{\Gamma(S\rightarrow D_s^{+}\pi^0)}$ with $S=D_{s0}^+(2317)$. 
Similar resonances which are degenerate with it have been observed 
in $B$ decays: 
$B\rightarrow\bar D\tilde D_{s0}^+(2317)[D_s\pi^0,
D_s^{*+}\gamma]$~\cite{BELLE-BD}, 
$B\rightarrow \bar D ({\rm or}\,\,\bar D^{*})
\tilde D_{s0}^+(2317)\break[D_s\pi^0]$~\cite{BABAR-B}. 
Here the new resonances have been denoted by 
$\tilde D_{s0}^+(2317)$[observed channel(s)] to distinguish them 
from the above $D_{s0}^+(2317)$, although they are usually 
identified to $D_{s0}^+(2317)$. It is because the resonance signals 
have been observed in the $D_s^{*+}\gamma$ channel in addition to 
the $D_{s}^+\pi^0$~\cite{BELLE-BD}. 
It is quite different from the previous $D_{s0}^+(2317)$. 

As will be seen later, assigning $D_{s0}^+(2317)$ to the $I_3=0$ 
component $\hat F_I^+$~\cite{Terasaki-D_s} of iso-triplet scalar 
four-quark mesons, 
$\hat F_I\sim [cn][\bar s\bar n]_{I=1},\,(n=u,d)$, 
is favored by Eq.~(\ref{eq:constraint}). In this case, 
its narrow width might be wondered because 
$\hat F_I^+\rightarrow D_s^+\pi^0$ appears a {\it fall-apart} decay 
at a glance. However, its small rate can be realized by a small 
overlap between  wavefunctions of the initial 
$|D_{s0}^+(2317)\rangle$ and final $\langle{D_s^+\pi^0}|$ states. 
Such a small overlap can be seen by decomposing a scalar four-quark 
state $|[qq][\bar q\bar q]\rangle$ with ${\bf 1_s}\times {\bf 1_s}$ 
of spin $SU(2)$ and ${\bf \bar 3_c}\times {\bf 3_c}$ of color 
$SU_c(3)$, which is the lowest lying four-quark state~\cite{Jaffe}, 
into a sum of $|\{q\bar q\}\{q\bar q\}\rangle$ states; 
\begin{eqnarray}
\hspace{-10mm}
 |[qq]_{\bf \bar 3_c}^{\bf 1_s}
[\bar q\bar q]_{\bf 3_c}^{\bf 1_s}\rangle    
&& =-\sqrt{1\over 4}\sqrt{1\over 3}
|\{q\bar q\}_{\bf 1_c}^{\bf 1_s}     
\{q\bar q\}\}_{\bf 1_c}^{\bf 1_s}\rangle \nonumber\\
&&\hspace{4mm} + \sqrt{3\over 4}\sqrt{1\over 3}
|\{q\bar q\}_{\bf 1_c}^{\bf 3_s}
\{q\bar q\}\}_{\bf 1_c}^{\bf 3_s}\rangle + \cdots.  
\label{eq:decomp}
\end{eqnarray}
The color and spin wavefunction overlap between 
$|\hat F_I^+\rangle$ and $\langle{D_s^+\pi^0}|$ is given by 
the first term of the right-hand-side of Eq.~(\ref{eq:decomp}). 
To see its narrow width more explicitly, we estimate the 
rate for $\hat F_I^+\rightarrow D_s^+\pi^0$ by comparing it with 
$\hat\delta^{s+}\rightarrow \eta\pi^+$. Here we have assigned the 
observed $a_0(980)$, $f_0(980)$, $\kappa(800)$ and 
$f_0(600)$~\cite{PDG04} to scalar $[qq][\bar q\bar q]$ mesons, 
$\hat\delta^s$, $\hat\sigma^s$, $\hat\kappa$ and 
$\hat\sigma$~\cite{Jaffe}. 
However, the above overlap can be 
quite different from that of $\langle{\eta\pi^+}|$ and 
$|{\hat\delta^{s+}}\rangle$ at the scale of $m_{\hat\delta^s}\sim 1$ 
GeV, because a gluon exchange between $\{q\bar q\}$ pairs will 
reshuffle the above decomposition [while such a reshuffling will be 
rare at the 2 GeV or a higher energy scale, because it is known that 
the $s$-quark at the 2 GeV scale is much more slim 
($m_s\simeq 90$ MeV)~\cite{Gupta} than the $s$-quark in the 
constituent quark model, i.e., the quark-gluon coupling at 2 GeV 
scale is much weaker than that at 1 GeV scale]. With this in mind, 
we introduce a parameter $\beta_0$ describing the difference 
between overlaps of color and spin wavefunctions at the scale of 
$m_{\hat\delta^s}$ and at the scale of $m_{\hat F_I^+}$.   
In the limiting case that the full reshuffling around 1 GeV 
while no reshuffling at the scale of $m_{\hat F_I}$, we have 
$|\beta_0|^2={1}/{12}$ as seen in Eq.~(\ref{eq:decomp}). 
By using a hard pion approximation and the asymptotic $SU_f(4)$ 
symmetry, which have been reviewed comprehensively in 
Ref.~\cite{suppl}, we have 
\begin{equation}
\Gamma(\hat F_I^+\rightarrow D_s^+\pi^0)_{SU_f(4)}\simeq 5 - 10 \,\, 
{\rm MeV}
\label{eq:width-SU_f(4)}
\end{equation} 
where the spatial wavefunction overlap is in the $SU_f(4)$ symmetry 
limit. Here, we have used 
$\Gamma(a_0(980)\rightarrow \eta\pi)_{\rm exp}= 50 - 100$ MeV and  
the $\eta$-$\eta'$ mixing angle 
$\theta_P\simeq -20^\circ$~\cite{PDG04} 
as the input data. Noting that the above $SU_f(4)$ 
symmetry overestimates by $20 - 30$ \% in amplitude, we have 
$\Gamma(\hat F_I^+)\simeq \Gamma(\hat F_I^+\rightarrow D_s^+\pi^0)
\sim 3.5 - 7$ MeV~\cite{HT-isospin}. 
It is sufficiently narrow.  

Next, we study the radiative decay of $D_{s0}^+(2317)$ to see that 
its assignment to $\hat F_I^+$ is consistent with 
Eq.~(\ref{eq:constraint}). For later convenience, we study the 
typical three cases, $D_{s0}^+(2317)$ as 
(i) the iso-triplet $\hat F_I^+$,  
(ii) the iso-singlet $\hat F_0^+\sim [cn][\bar s\bar n]_{I=0}$ and 
(iii) the conventional scalar $D_{s0}^{*+}\sim \{c\bar s\}$,  
under the vector meson dominance (VMD) hypothesis. To test our 
approach, we study $D_s^{*+}\rightarrow D_s^+\gamma$ in the same 
way. Here, we take $VVP$ and $SVV$ couplings with spatial 
wavefunction overlap in the $SU_f(4)$ symmetry limit, where $V$, $P$ 
and $S$ denote a vector, a pseudoscalar and a scalar meson, 
respectively, and take the overlapping factor $|\beta_1|^2={1}/{4}$ 
between wavefunctions of a scalar four-quark and two vector-meson 
states, as seen in Eq.~(\ref{eq:decomp}). The results are listed in 
Table~\ref{tab:1}, where the input data are taken from 
Ref.~\cite{PDG04}. Comparing the rate for 
$\hat F_I^+\rightarrow D_s^{*+}\gamma$ in Table~\ref{tab:1} with 
Eq.~(\ref{eq:width-SU_f(4)}), we obtain 
$R(\hat F_I^+) \sim (4.5 - 9)\times 10^{-3}$~\cite{HT-isospin},        
which is consistent with the constraint Eq.~(\ref{eq:constraint}). 
It implies that assigning $D_{s0}^+(2317)$ to $\hat F_I^+$ is 
favored by experiments. 
\begin{table}[t]
\caption{Rates for radiative decays of charm-strange mesons under 
the VMD, where  spatial wavefunction overlap is in the $SU_f(4)$ 
symmetry. The input data are taken from Ref.~\cite{PDG04}. }
\label{tab:1}       
\begin{tabular}
{c c c c}
\hline\noalign{\smallskip}
Decay & Pole(s) 
& Input Data & 
Rate (keV) \\
\noalign{\smallskip}\hline\noalign{\smallskip}
$D_s^{*+}\rightarrow D_s^+\gamma$ & 
$\phi,\,\psi$ 
& $\Gamma(\omega\rightarrow \pi^0\gamma)_{\rm exp}$ & 
{0.8}\\
\noalign{\smallskip}\hline\noalign{\smallskip}
{$\hat F_I^+\rightarrow D_s^{*+}\gamma$} & 
{$\rho^0$} 
& $\Gamma(\phi\rightarrow a_0\gamma)_{\rm exp}$ & 
{45}\\
\noalign{\smallskip}\hline\noalign{\smallskip}
$\hat F_0^+\rightarrow D_s^{*+}\gamma$ & 
{$\omega$} 
& $\Gamma(\phi\rightarrow a_0\gamma)_{\rm exp}$ & 
{4.7} \\
\noalign{\smallskip}\hline\noalign{\smallskip}
{$D_{s0}^{*+}\rightarrow D_s^{*+}\gamma$} & 
{$\phi,\,\psi$} 
& $\Gamma(\chi_{c0}\rightarrow \psi\gamma)_{\rm exp}$ & 
{35}\\
\noalign{\smallskip}\hline
\end{tabular}
\end{table}

In the cases of the above assignments (ii) and (iii),  
$D_{s0}^+(2317)\rightarrow D_s^+\pi^0$ is isospin non-conserving. 
The isospin non-conservation is assumed to be caused by the 
$\eta$-$\pi^0$ mixing as usual. The mixing parameter $\epsilon$ has 
been estimated~\cite{Dalitz} as 
$\epsilon = 0.0105\pm 0.0013$, 
i.e., $\epsilon \sim O(\alpha)$ with the fine structure constant 
$\alpha$. It implies that isospin non-conserving decays are much 
weaker than the radiative ones. 
By using the hard pion approximation, the asymptotic $SU_f(4)$ 
symmetry and  the above value of $\epsilon$, the rates for the 
isospin non-conserving decays can be obtained as listed in 
Table~\ref{tab:2}.  
\begin{table}[b]
\caption{Rates for isospin non-conserving decays, where the spatial 
wavefunction overlap is in the $SU_f(4)$ symmetry. Input data are 
taken from Ref.~\cite{PDG04}. 
}
\label{tab:2}       
\begin{tabular}
{c l c}
\hline\noalign{\smallskip}
Decay & Input Data 
& Rate (keV)  \\
\noalign{\smallskip}\hline\noalign{\smallskip}
{$D_s^{*+}\rightarrow D_s^+\pi^0$} 
& {$\Gamma(\rho\rightarrow\pi\pi)_{\rm exp}$} 
& {0.05} \\
\noalign{\smallskip}\hline\noalign{\smallskip}
{$\hat F_0^+\rightarrow D_s^+\pi^0$} 
& {$\Gamma(a_0\rightarrow \eta\pi)\simeq 70$ MeV} 
& {0.7} \\
\noalign{\smallskip}\hline\noalign{\smallskip}
{$D_{s0}^{*+}\rightarrow D_s^+\pi^0$} 
& {$\Gamma(K_0^{*0} \rightarrow K^+\pi^-)_{\rm exp}$} 
& {0.6}
\\
\noalign{\smallskip}
\hline  
\end{tabular}
\end{table}
The results on the decays of $D_s^{*+}$ in Tables~\ref{tab:1} 
and \ref{tab:2} lead to the ratio of decay rates 
$R(D_s^{*+})^{-1}\simeq 0.06$. This reproduces well the experimental   
value~\cite{BABAR-D_s^*} 
$R(D_s^{*+})^{-1}_{\rm BABAR}=0.062\pm 0.005\pm 0.006$. 
This means that the present approach is sufficiently reliable.  
The corresponding ratios of the decay rates 
in the cases (ii) and (iii) are also obtained as   
(ii) $R(\hat F_0^+)\simeq 7$ and 
(iii) $R(D_{s0}^{*+})\simeq 60$.   
They are much larger than the experimental upper bound. It should 
be noted that the isospin non-conserving decays are much weaker 
than the radiative decays, as expected intuitively above. 
The assignment of $D_{s0}^+(2317)$ to the iso-singlet $DK$ 
molecule~\cite{BCL} leads to $R(\{DK\})\simeq 3$ which is 
much larger than the experimental upper bound in 
Eq.~(\ref{eq:constraint}). Hence, such an assignment should be 
rejected~\cite{MS}. Thus, assigning $D_{s0}^+(2317)$ to an 
iso-singlet state ($\hat F_0^+$, $D_{s0}^{*+}$ or $DK$ molecule) is 
disfavored by experiments. 

From the above considerations, it is natural to assign 
$D_{s0}^+(2317)$ to the iso-triplet $\hat F_I^+$. However, its 
neutral and doubly charged partners, $\hat F_I^{++}$ and 
$\hat F_I^0$, have not yet been observed~\cite{BABAR-search}. 
With this in mind, we study the production of charm-strange scalar 
mesons ($\hat F_I^{++,+,0}$ and $\hat F_0^+$) 
by assigning $D_{s0}^+(2317)$ to $\hat F_I^{+}$, and discuss 
why experiments have observed $D_{s0}^+(2317)$ but not its neutral 
and doubly charged partners. To this aim, we consider their 
production through weak interactions as a possible mechanism, 
because the OZI-rule violating productions of multi-$q\bar q$-pairs 
and their recombinations into four-quark meson states are believed 
to be strongly suppressed at high energies. First, we recall that 
color mismatched decays which include rearrangements of colors in 
weak decay processes would be suppressed compared with color favored 
ones as long as non-factorizable contributions, which are actually 
small in $B$ decays and are expected to be much smaller at higher 
energies, are ignored. Next, we draw quark-line diagrams within the 
minimal $q\bar q$ pair creation, noting the OZI rule. 
\begin{figure}[!t]  %
\includegraphics[width=75mm,clip]{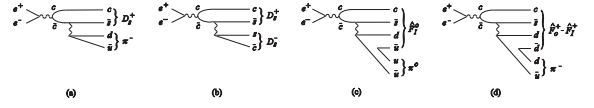}
\label{fig:via-c-cbar}       %
\caption{
Production of charm-strange scalar four-quark mesons 
through $e^+e^-\rightarrow c\bar c$. (a) and (b) describe the 
production $D_s^+\pi^-,\,D_s^{*+}\pi^-$, etc., and 
$D_s^+\pi^-,\,D_s^{*+}\pi^-$, etc., respectively. The production of 
$\hat F_I^0$, $\hat F_I^+$ and $\hat F_0^+$ is given by (c) and (d). 
}
\end{figure}%
Because there is no diagram yielding $\hat F_I^{++}$ production 
in this approximation, as seen in Fig.~1, it is easy to understand why 
no evidence of $\hat F_I^{++}$ was found in 
$e^+e^-\rightarrow c\bar c$ experiments. As will be seen in 
productions of $\tilde D_{s0}^+(2317)[D_s^+\pi^0]$ in the $B$ 
decays, their observed rates are comparable with color mismatched 
decays, so that rates for $\hat F_I^0$ and $\hat F_0^+$ productions 
in $e^+e^-\rightarrow c\bar c$ annihilation will be expected to be 
comparable with that through the color suppressed ones. 
Therefore, they would be much weaker (possibly by about two order 
of magnitude) than productions of 
$D_s^+\pi^-,\,D_s^{*+}\pi^-,\,D_s^{*+}\rho^- $, etc.,  
and $D_s^+D_s^-,\,D_s^{*+}D_s^-,\,D_s^{*+}D_s^{*-}$, etc.,  
created through the reaction depicted by Figs.~1(a) and (b). 
The $D_s^+\pi^-$ produced through Fig.~1(a) obscures the signal 
$\hat F_I^0\rightarrow D_s^+\pi^-$ events. In addition, the  
$D_s^{*+}$ and $\gamma$ from $D_s^{*-}\rightarrow D_s^-\gamma$ 
produced through the ordinary 
$e^+e^-\rightarrow c\bar c\rightarrow D_s^{*+}D_s^{*-}$ 
[and through Fig.~1(b)] obscure the signal of 
$\hat F_0^+\rightarrow D_s^{*+}\gamma$ events. 
Therefore, it is understood why the inclusive $e^+e^-$ 
annihilation experiment found no signal of scalar resonance in the 
$D_s^+\pi^-$ and $D_s^{*+}\gamma$ channels. In the case of 
$\hat F_I^+$, however, there do not exist large numbers 
of background events described by Figs.~1(a) and (b), because its 
main decay is $\hat F_I^+\rightarrow D_s^+\pi^0$. In fact, 
$D_{s0}^+(2317)$ has been observed in the $D_s^+\pi^0$ channel. 
This seems to imply that the production of four-quark mesons in 
hadronic weak decays plays an essential role~\cite{production}. 
\begin{figure}[!h]     %
\begin{center}  
\includegraphics[width=70mm,clip]{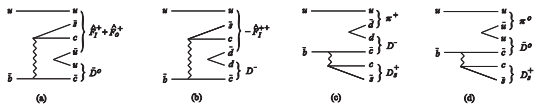}   
\label{fig:Bu-8.eps}              %
\caption{Production of charm-strange scalar mesons 
in the $B_u^+$ decays. (c) and (d) describe the production of 
backgrounds of $\hat F_I^{++}$ and $\hat F_I^+$ signals, 
respectively. 
}
\end{center} 
\end{figure}%

Because it is difficult to observe $\hat F_I^{++}$ and 
$\hat F_I^{0}$ in inclusive $e^+e^-\rightarrow c\bar c$ experiments, 
we study their productions in $B$ decays. First, we draw quark-line 
diagrams in the same way as the above. As expected in Figs.~2(a) and 
3(b), $\tilde D_{s0}^+(2317)[D_s^+\pi^0]$, which can be identified 
to $\hat F_I^+$ because it decays dominantly into $D_s^+\pi^0$ as 
seen above, has been observed. The production of $\hat F_I^{++}$ is 
given by Fig.~2(b) which is of the same type as Fig.~2(a). In 
addition, the production of $\hat F_I^0$ is given by Fig.~3(a) which 
is again of the same type as Figs.~2(a) and (b), so that their 
production rates are not very different from each other; 
$B(B_u^+\rightarrow \hat F_I^{++}D^-) 
\sim B(B_d^+\rightarrow \hat F_I^{0}\bar D^0)
\sim  B(B_u^+\rightarrow \hat F_I^{+}\bar D^0)$, 
where 
$B(B_u^+\rightarrow \hat F_I^{+}\bar D^0)_{\rm exp}
\sim 10^{-3}$~\cite{BELLE-BD,BABAR-B}. 
Besides, the BELLE observed indications of 
$\tilde D_{s0}^+(2317)[D_s^{*+}\gamma]$ which are conjectured to be 
signals of $\hat F_0^+\rightarrow D_s^{*+}\gamma$ 
because the production of $\hat F_0^+$ and $\hat F_I^+$ are depicted 
by the same diagrams and the $\hat F_I^+\rightarrow D_s^{*+}\gamma$ 
is much weaker than the $\hat F_I^+\rightarrow D_s^+\pi^0$ while the 
$\hat F_0^+\rightarrow D_s^{*+}\gamma$ is much stronger than the 
$\hat F_0^+\rightarrow D_s^{+}\pi^0$, as seen before. 
\begin{figure}[!t]    %
\includegraphics[width=70mm,clip]{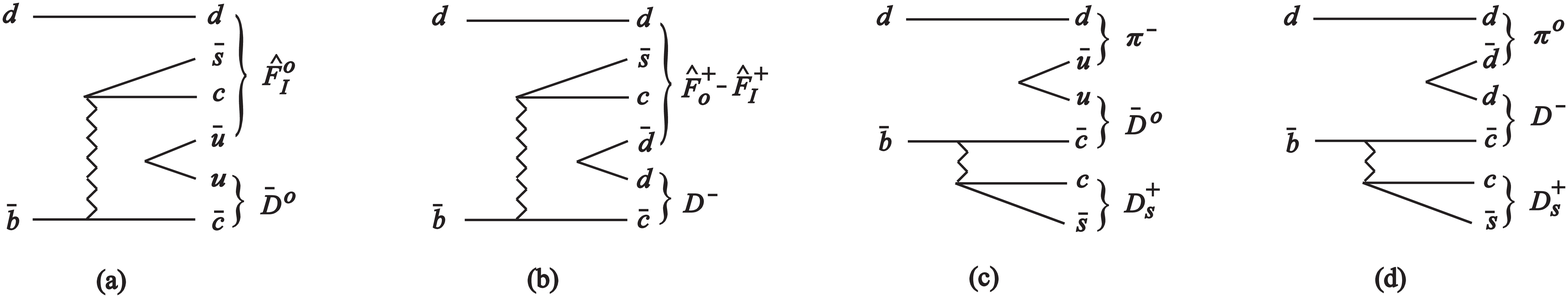}   
\label{fig:Bd-8.eps}             %
\caption{ Production of $\hat F_I^+$, $\hat F_0^+$ and 
$\hat F_I^{0}$ in the $B_d^0$ decays. (c) and (d) depict the 
production of  backgrounds of $\hat F_I^{0}$ and $\hat F_I^+$. 
}
\end{figure}%
As expected in Fig.~4(c), the BELLE~\cite{BELLE-BK} observed 
$\bar B_d\rightarrow \tilde D_{s0}^+(2317)[D_s^+\pi^0]K^-$, 
and provided 
$B(\bar B_d\rightarrow \tilde D_{s0}^+(2317)K^-)
\cdot B(\tilde D_{s0}^+(2317)\rightarrow D_s^+\pi^0)
=(5.3^{+1.5}_{-1.3}\pm 0.7\pm 1.4)\times 10^{-5}$. 
If $\tilde D_{s0}^+(2317)[D_s^+\pi^0]$ is identified to 
$\hat F_I^+$ and 
$B(\tilde D_{s0}^+(2317)\rightarrow D_s^+\pi^0)\simeq 100\,\%$ 
is taken, 
$B(\bar B_d^0\rightarrow \hat F_I^+K^-)\sim 10^{-5}-10^{-4}$ 
would be obtained. 
Using it as the input data and noting that Figs.~4(c) and 5(c) are 
of the same type, we could estimate 
$B(B_u^-\rightarrow K^-\hat F_I^0)\sim 10^{-5}-10^{-4}$, 
if contributions from diagrams Figs.~5(a) and (b) cancel each
other (because the phases of $\hat F_I^0$ in these diagrams have 
opposite signs arising from anti-symmetry property of its 
wavefunction). 
\begin{figure}[!t]    %
\includegraphics[width=70mm,clip]{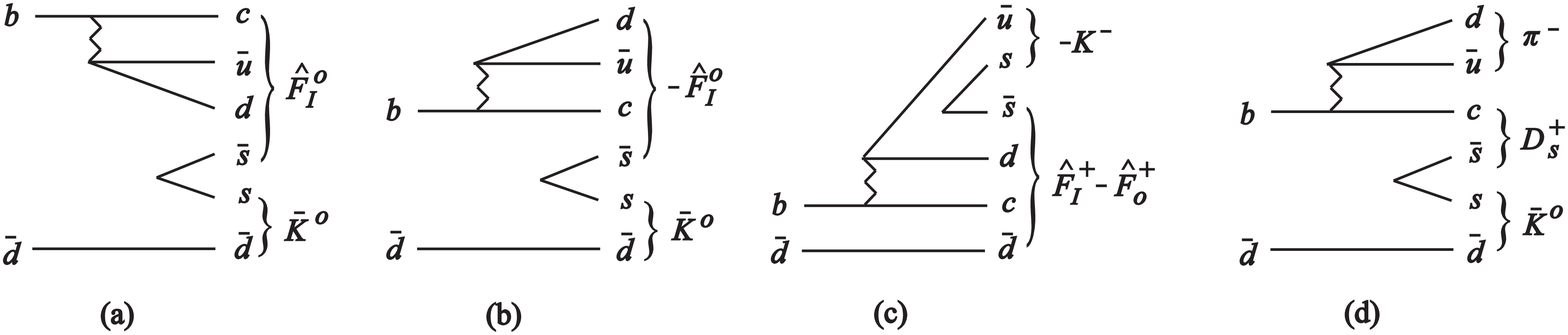}       
\label{fig:barBu-8.eps}     %
\caption{Production of $\hat F_I^+$, $\hat F_I^0$ and $\hat F_0^+$ 
in the decays of $\bar B_d^0$. (d) describes the production of 
backgrounds of $\hat F_I^0$ signal.}
\end{figure}%
\begin{figure}[!b]    %
\includegraphics[width=70mm,clip]{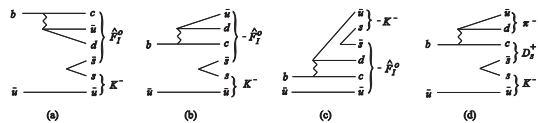}     
\label{fig:barBd-8.eps}                %
\caption{Production of $\hat F_I^0$ in the $B_u^-$ decays. 
(d) describes the production of backgrounds of its signals.}
\end{figure}%

In summary, we have seen that assigning $D_{s0}^+(2317)$ to an 
iso-triplet $\hat F_I^+$ is favored by experiments. 
In addition, we have discussed why inclusive 
$e^+e^-\rightarrow c\bar c$ experiments observed no evidence for 
its neutral and doubly charged partners $\hat F_I^0$ and 
$\hat F_I^{++}$. $\tilde D_{s0}^+(2317)[D_s^+\pi^0]$ which was 
observed in $B$ decays has been identified to $D_{s0}^+(2317)$. 
Indications of $\hat F_0^+$ also have been observed as 
$\tilde D_{s0}^+(2317)[D_s^{*+}\gamma]$ in $B$ decays. $\hat F_I^0$ 
and $\hat F_I^{++}$ will be observed in $B$ decays. \vspace{-10mm}\\
\section*{Acknowledgments} 
The author would like to thank the organizers of QNP06 for 
supporting local expense during the conference. 
This work is supported in part by the Grant-in-Aid for Science 
Research, Ministry of Education, Culture, Sports, Science and 
Technology of Japan (No.~16540243). 

\end{document}